\documentclass[conference]{IEEEtran}
\usepackage{graphicx}  
\usepackage{booktabs}        
\usepackage{amsmath}
\usepackage{amssymb}
\usepackage{siunitx}          \usepackage{url}
\usepackage{array}       
\usepackage{multirow}
\usepackage{tabularx}               \usepackage[caption=false,font=footnotesize]{subfig}
\usepackage[hidelinks]{hyperref}
\usepackage{threeparttable}
\usepackage{float}
\usepackage{stfloats}
\usepackage{textcomp}
\IEEEoverridecommandlockouts                   
                       
\title{\LARGE \bf
VISTA: A Benchmark for Real-Time Video Streaming under Network Impairments in Surgical Teleoperation}

\author{Zexin Deng$^{*}$, Zhenhui Yuan$^{*}$, Tian Lu$^{*}$, Gaofeng Li$^\dagger$, Meipeng Huang$^\ddagger$, and Longhao Zou$^\S$%
\thanks{$^{*}$Z. Deng, Z. Yuan, and T. Lu are with the School of
Engineering, University of Warwick, Coventry CV4 7AL, United Kingdom
(e-mail: \{zexin.deng, zhenhui.yuan, lu.tian.2\}@warwick.ac.uk). Z.~Yuan is a consultant member at the Institute for Applied and Translational Technologies in Surgery, University Hospitals Coventry \& Warwickshire NHS Trust, UK.}
\thanks{$^\dagger$G. Li is with the College of Control Science and
Engineering and the State Key Laboratory of Industrial Control Technology,
Zhejiang University, China (e-mail: gaofeng.li@zju.edu.cn).}%
\thanks{$^\ddagger$M. Huang is with Pengcheng Laboratory, Shenzhen 518000, China, and also with South China University of Technology, Guangzhou 511442, China (e-mail: huangmp@pcl.ac.cn).}%
\thanks{$^\S$L. Zou is with Pengcheng Laboratory, Shenzhen 518000, China, and also with Southern University of Science and Technology, China (e-mail: zoulh@pcl.ac.cn).}%
}
\begin{document}
\maketitle
\thispagestyle{empty}
\pagestyle{empty}

\begin{abstract}
Real-time video streaming is crucial in surgical teleoperation, yet reproducible evaluation under realistic network impairments remains limited. This paper presents \textit{VISTA}, a benchmark designed to study how impairments along the forward video path affect received video quality, temporal continuity, and human task performance. \textit{VISTA} employs Linux Traffic Control with NetEm and a Gilbert-Elliott loss model to emulate five network conditions: Hospital LAN, 5G Urban, 4G Rural, LEO Satellite, and GEO Satellite. The benchmark integrates a standardised peg transfer task with
synchronized measurements of network quality of service (QoS),
objective video quality (PSNR, SSIM, and VMAF), and temporal
continuity through freeze rate, while maintaining a stable reverse
control channel. Across 375 experimental trials, network degradation substantially reduced teleoperation performance: success rate decreased from 97\% in Hospital LAN to 79\% in 5G Urban, 35\% in 4G Rural, 71\% in LEO Satellite, and 12\% in GEO Satellite, while mean task completion time for successful trials increased from 80 s in Hospital LAN to 117 s in 5G Urban, 211 s in 4G Rural, 152 s in LEO Satellite, and 255 s in GEO Satellite. These findings show that network impairments have a direct impact on task completion and success in surgical teleoperation, and provide a reproducible basis for evaluating teleoperation video under realistic network constraints. Source code available: \url{https://github.com/Dzxx623/VISTA}.
\end{abstract}


\section{INTRODUCTION}

Surgical teleoperation has the potential to expand access to specialist surgical care. The clinical need remains substantial, with the Lancet Commission on Global Surgery estimating that around 5~billion people lack access to safe and affordable surgical care~\cite{Meara2015}. Robot-assisted telesurgery has progressed from landmark demonstrations such as the Lindbergh operation toward renewed clinical and translational interest~\cite{marescaux2001transatlantic,rocco2024insights}. 

\begin{figure}[!t]
    \centering
    \includegraphics[
        width=\linewidth,
        trim=1.5cm 15cm 1.5cm 1.0cm,
        clip
    ]{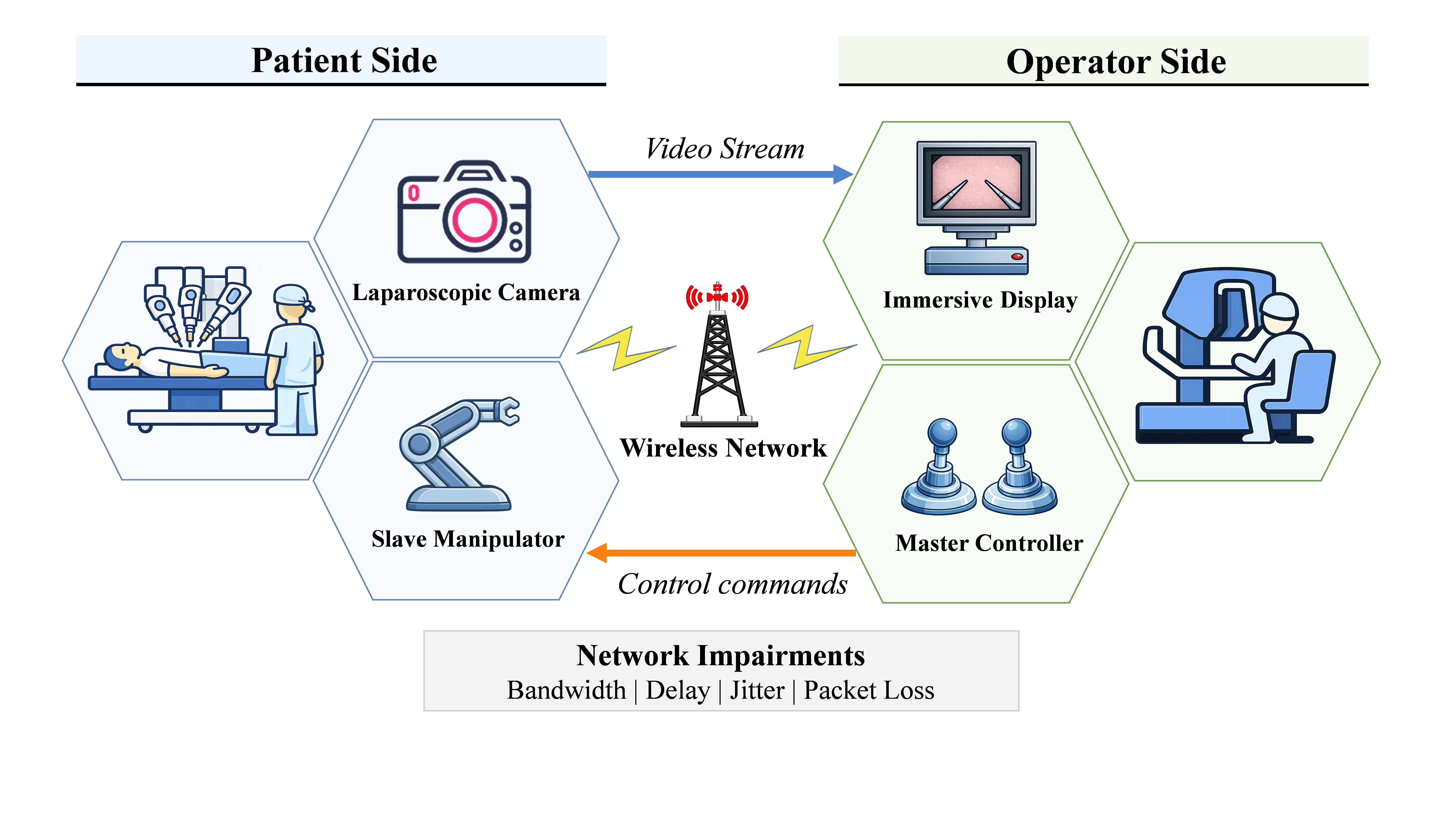}
    \vspace{-6mm}
    \caption{Communication asymmetry in surgical teleoperation, with laparoscopic video transmitted through the forward video path and motion commands through the reverse control path.}
    \label{fig:intro_video_impairment}
\end{figure}

However, the practical deployment of these systems faces a central communication challenge, driven primarily by the asymmetry between their two data paths, as illustrated in Fig.~\ref{fig:intro_video_impairment}. For human-in-the-loop operation, system usability depends critically on the fidelity, continuity, and timeliness of the forward video stream~\cite{akasaka2022impact}. This stringency is also reflected in recent 3GPP requirements for remote surgical video, which target end-to-end latency below 20\,ms and communication service availability above 99.9999\%~\cite{3GPP22263}. When latency, jitter, and packet loss increase, or available bandwidth becomes constrained, the surgeon may experience spatial distortion, temporal freezing, or stale scene updates, which in turn degrade hand--eye coordination and task execution~\cite{wang2025influence,Motiwala2025}. In this setting, degraded video is more than a quality of service (QoS) concern because it directly affects teleoperation performance and safety.

While prior telesurgery studies and related work on remote video operation indicate that adverse communication conditions and temporal disruptions can impair operator performance~\cite{akasaka2022impact, wang2025influence, Baltaci2022}, reproducible evaluation under controlled and comparable network impairments remains limited. Existing methodologies are difficult to compare because they often vary only a limited subset of network factors, focus separately on communication QoS or task performance, or rely on computer-based simulation and experimental settings outside surgical teleoperation~\cite{Deng2025TeleSim}. As summarized in Table~\ref{tab:related_work_comparison}, an open benchmark that jointly provides isolation of the forward video path, scenario-based multi-factor impairment profiles, and aligned measurements of network QoS, video quality, temporal continuity, and task performance remains lacking. \textit{VISTA} is designed to address this gap through a publicly released benchmark.

\begin{table*}[!t] 
  \centering
  \caption{Comparison with representative prior work in surgical teleoperation.}
  \label{tab:related_work_comparison}
  \renewcommand{\arraystretch}{1.4}
  \setlength{\tabcolsep}{4pt}
  \resizebox{\textwidth}{!}{%
  \begin{tabular}{|l|c|c|c|c|c|}
    \hline
    \textbf{Work} &
    \textbf{Platform} &
    \textbf{Network Impairment} &
    \textbf{Video Quality Evaluation} &
    \textbf{Task Performance} &
    \textbf{Reproducibility} \\
    \hline
    Xu et al.~\cite{Xu2014}
    & Simulator (dV-Trainer)
    & Latency
    & Subjective 
    & Time, Errors
    & Closed \\
    \hline
    Kumcu et al.~\cite{Kumcu2017}
    & Manual box (Non-robotic)
    & Latency
    & None
    & Time, Usability
    & Closed \\
    \hline
    Akasaka et al.~\cite{akasaka2022impact}
    & HIL (Proprietary robot)
    & Bandwidth, Packet loss
    & Subjective 
    & Time, Fatigue
    & Closed \\
    \hline
    Wang et al.~\cite{wang2025influence}
    & HIL (Proprietary robot)
    & Bandwidth, Latency
    & Subjective 
    & Time, Workload
    & Closed \\
    \hline
    \textbf{VISTA (ours)}$^\ast$
    & \textbf{HIL (Generic with Adapter)}
    & \textbf{Multi-factor}
    & \textbf{Objective VQA}
    & \textbf{Time, Success Rate, Failure Analysis}
    & \textbf{Open (Data + CAD)} \\
    \hline
  \end{tabular}%
  }
  
  \vspace{4pt}
  \parbox{\textwidth}{\scriptsize
    $^\ast$\textit{VISTA} employs a \textbf{universal mechanical adapter}, with the original Creo CAD files released in this work, to reduce dependence on proprietary hardware platforms. It further evaluates \textbf{multi-factor network impairments} across five realistic tiers and uses \textbf{objective VQA} (PSNR, SSIM, VMAF) instead of subjective rating scales.
  }
\end{table*}

To address this gap, we present \textit{VISTA} (\textbf{V}ideo
\textbf{I}mpairment \textbf{S}cenario \textbf{T}estbed for Surgical
Teleoperation \textbf{A}ssessment), a hardware-in-the-loop (HIL)
benchmark for evaluating video feedback under network impairments in
robot-assisted laparoscopic teleoperation. A key design choice in
\textit{VISTA} is to impair only the forward video path while keeping
the reverse control path stable. This separation is important because
communication degradation in the control path can independently affect
robot motion and confound the interpretation of task failures; prior
telesurgery work has also shown that delay and asynchrony in video and
control feedback can affect surgical task performance~\cite{Thompson1999}.
By holding the control path constant, \textit{VISTA} provides a
controlled basis for analysing how network impairments in the video
stream propagate to received video quality, temporal continuity, and
human task performance. The benchmark uses a standardised teleoperation
task and records aligned measurements of configured impairments,
measured QoS, received video quality, temporal continuity, and task
outcome. Our main contributions are as follows:

\begin{enumerate}
\item \textbf{A benchmark for video streaming performance under network impairments:}
We develop a controlled network emulation platform based on Linux
\textit{tc-netem} and virtual Ethernet pairs, and define five benchmark tiers: Hospital LAN, 5G Urban, 4G Rural, LEO Satellite, and GEO Satellite, which jointly model bandwidth, latency, jitter, and bursty packet loss using a Gilbert--Elliott model. Across 375 trials of an FLS-based teleoperation task, we identify three distinct failure modes: visual degradation in 4G Rural scenarios, persistent delay-induced temporal mismatch in GEO 
Satellite conditions, and episodic temporal instability in LEO Satellite environments. The benchmark protocol and dataset are released to support reproducible research on network-aware teleoperation systems.

\item \textbf{An integrated evaluation pipeline:}
\textit{VISTA} provides a unified measurement framework that captures network QoS, objective video quality (PSNR,
SSIM, VMAF), as well as temporal continuity indicators including one-way delay and freeze rate. It complements these
with end-to-end task performance measurements from
human-in-the-loop teleoperation trials. By aligning these heterogeneous metrics within a single evaluation pipeline, \textit{VISTA} enables a consistent and reproducible analysis of how network impairments propagate through the communication stack and ultimately affect human teleoperation performance.
\end{enumerate}

\section{HARDWARE-IN-THE-LOOP TESTBED AND BENCHMARK DESIGN}

This section details the \textit{VISTA} testbed used to isolate the
forward video path, together with the network impairment injection
mechanism and the design of the benchmark network tiers.

\subsection{System Overview}

\begin{figure*}[!t]
  \centering
  \vspace{-2mm}
  \includegraphics[
    width=0.94\textwidth,
    trim=2cm 25cm 4cm 18cm,
    clip
  ]{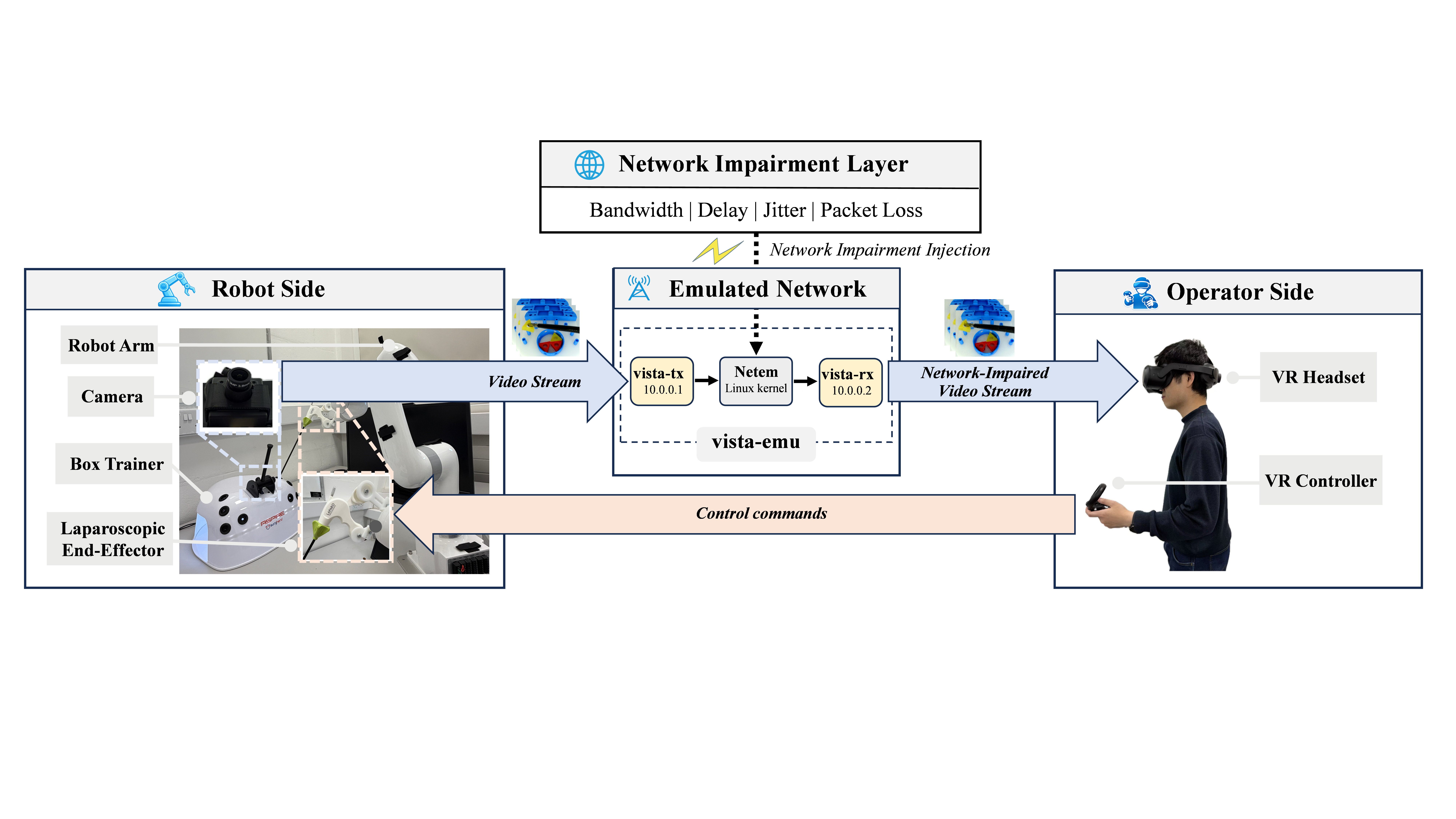}
  \vspace{-9mm}
  \caption{Overview of the \textit{VISTA} HIL testbed at the University of Warwick. The emulated network injects impairments on the forward video path between the laparoscopic box trainer and the operator side, while control commands are transmitted to the UFactory Lite~6 manipulator over a dedicated unimpaired low-latency link.}
  \label{fig:physical_setup}
  \vspace{-1mm}
\end{figure*}

The overall \textit{VISTA} platform is illustrated in Fig.~\ref{fig:physical_setup}. The surgical workspace is provided by an ASPIRE Surghero laparoscopic
box trainer, whose integrated endoscopic camera captures RGB video of
the manipulation scene and streams it to a workstation equipped with an
NVIDIA RTX 4060 GPU for packet reception, packet assembly, decoding,
visualisation, and metric logging. The decoded stream is rendered
through a fixed VIVE Focus\,3 headset pipeline. On the host PC, HTC
VIVE Business Streaming is used, with HTC VIVE Hub installed, to
deliver SteamVR content to the headset via a USB 3.0 cable. This
pipeline remains unchanged across all trials. The operator issues motion commands through a VR controller connected
to the same workstation, and these commands are forwarded over a
dedicated low-latency Gigabit Ethernet link to a six-degree-of-freedom
(6-DOF) UFactory Lite~6 manipulator. This separation between the impaired forward video path and the
stable reverse control path ensures that robot actuation is unaffected
by degradation on the video path. Controller pose and button state are
sampled at 120\,Hz through the OpenVR API.

\subsection{Network Impairment Injection}

Controlled impairments are applied only to the forward video path using
an HIL pipeline implemented on a single workstation. Video packets are
sent through a Linux virtual Ethernet (veth) pair connected to an
isolated network namespace, where the impaired stream is relayed back
for local decoding and rendering; the reverse control path is kept
unimpaired.

Impairments are configured with Linux \texttt{tc} using NetEm as the
root queueing discipline on the host-side veth interface
(\texttt{veth-host})~\cite{tcnetem}. For each trial, delay, jitter,
packet loss, and bandwidth are applied atomically in a single NetEm
rule. Bandwidth shaping uses NetEm's native \texttt{rate} parameter
rather than a separate token bucket filter, avoiding interaction between
multiple queueing disciplines.

\subsection{Network Tier Configuration}

\textit{VISTA} defines five benchmark network tiers on the forward video
path: Hospital LAN, 5G Urban, 4G Rural, LEO Satellite, and GEO
Satellite. Each tier jointly specifies bandwidth, nominal one-way delay,
jitter, and packet loss, spanning conditions from local connectivity to
links with larger propagation delay and stronger loss variation.

Packet loss is modeled using the Gilbert model implemented in NetEm
\texttt{gemodel}~\cite{tcnetem,gilbert1960}. In the simplified form
used here, the Good state is lossless and the Bad state is fully lossy,
allowing each tier to be specified by the target steady-state loss rate
and mean burst length:
\begin{equation}
  p_E = \frac{p}{p+r}, \qquad L_B = \frac{1}{r},
  \label{eq:ge}
\end{equation}
where $p$ and $r$ denote the Good$\to$Bad and Bad$\to$Good transition
probabilities, respectively~\cite{gilbert1960,hasslinger2008}. For each
tier, $p$ and $r$ are derived from the target $p_E$ and $L_B$ values
reported in Table~\ref{tab:nc_levels}.

The tier parameters were chosen to span representative teleoperation
conditions and were grounded in standard performance objectives,
published measurements, and clinical latency constraints. Hospital LAN
serves as a dedicated fibre baseline within ITU-T Y.1541
objectives~\cite{ITUY1541}. 5G Urban follows 3GPP TS~22.261 service
requirements and recent telesurgery studies~\cite{rocco2024insights,3GPP22261}.
4G Rural represents a degraded cellular setting near the operating
margin identified in~\cite{wang2025influence}. LEO Satellite reflects a
Starlink-class low Earth orbit link, consistent with reported latency
and packet loss behaviour~\cite{richter2025starlink}. GEO Satellite is
included as a high-latency reference condition that exceeds the
320\,ms clinical ceiling reported in~\cite{wang2025influence} and the
broader latency guidance in~\cite{ITUG114}.

\begin{table*}[!t]
  \centering
  \caption{Benchmark network tiers and impairment parameters.}
  \label{tab:nc_levels}
  \renewcommand{\arraystretch}{1.25}
  \setlength{\tabcolsep}{3.5pt}

  {\footnotesize
  \begin{tabular}{|l|c|c|c|c|c|c|c|}
    \hline
    \multirow{2}{*}{\textbf{Networks}} &
    \multirow{2}{*}{\textbf{Bandwidth (Mbps)}} &
    \multirow{2}{*}{\textbf{Nominal one-way delay (ms)}} &
    \multirow{2}{*}{\textbf{Jitter (ms)}} &
    \multicolumn{4}{c|}{\textbf{Packet loss}} \\
    \cline{5-8}
    & & & &
    \boldmath$p$ \textbf{(\%)} &
    \boldmath$r$ \textbf{(\%)} &
    \boldmath$p_E$ \textbf{(\%)} &
    \boldmath$L_B$ \textbf{(pkts)} \\
    \hline
    Hospital LAN  & 100 & 3   & 1  & 0.005 & 50.0 & 0.01 & 2 \\
    \hline
    5G Urban      & 20  & 35  & 8  & 0.167 & 33.3 & 0.50 & 3 \\
    \hline
    4G Rural      & 5   & 150 & 30 & 0.341 & 16.7 & 2.00 & 6 \\
    \hline
    LEO Satellite & 30  & 30  & 10 & 0.305 & 20.0 & 1.50 & 5 \\
    \hline
    GEO Satellite & 10  & 600 & 50 & 0.251 & 50.0 & 0.50 & 2 \\
    \hline
  \end{tabular}
  }

  \vspace{4pt}
  \parbox{0.9\textwidth}{\scriptsize
    \emph{Note:} One-way delay denotes the nominal delay configured in
    \texttt{tc~netem}; jitter denotes normally distributed delay
    variation around this value. $p$ and $r$ are the Good$\to$Bad and
    Bad$\to$Good transition probabilities passed to \texttt{gemodel}.
    $p_E$ is the steady-state loss rate, and $L_B$ is the mean burst
    length. $p$, $r$, and $p_E$ are reported in percent, whereas
    Eq.~\eqref{eq:ge} uses probability-scale values.
  }
\end{table*}

\section{EXPERIMENTAL PROTOCOL AND METRICS}

This section specifies the experimental protocol used to evaluate
\textit{VISTA}, including the task definition, video transport
configuration, and the trial-level measurement pipeline. Each trial
produces an aligned record of configured impairments, measured QoS,
received video quality, temporal continuity, and task outcome.

\subsection{Task Definition and Operator Preparation}

We adopt a Central-to-Peripheral Peg Transfer (C2P) task derived
from the Fundamentals of Laparoscopic Surgery (FLS) peg transfer
task. Unlike the canonical bimanual FLS task, the protocol is performed using a teleoperated \textit{UFactory Lite~6} manipulator under a fixed laparoscopic camera view, preserving the core demands of visually guided grasping, transfer, alignment, and release.

\begin{figure}[!t]
  \centering
  \subfloat[Workspace definition]{%
    \includegraphics[width=0.33\linewidth]{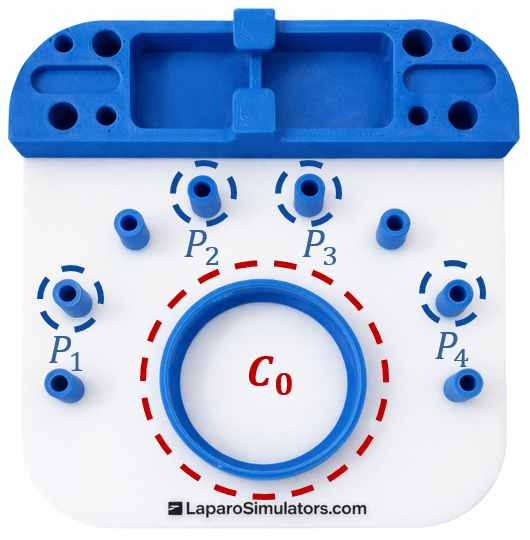}%
  }\hfill
  \subfloat[Initial state]{%
    \includegraphics[width=0.32\linewidth]{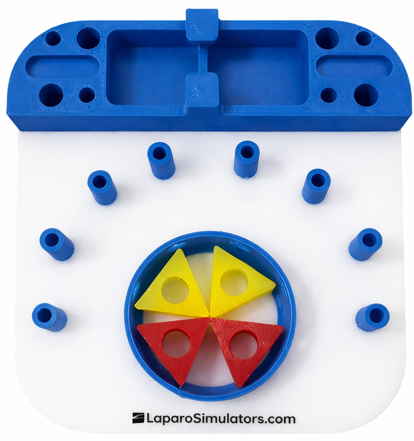}%
  }\hfill
  \subfloat[Target placement]{%
    \includegraphics[width=0.32\linewidth]{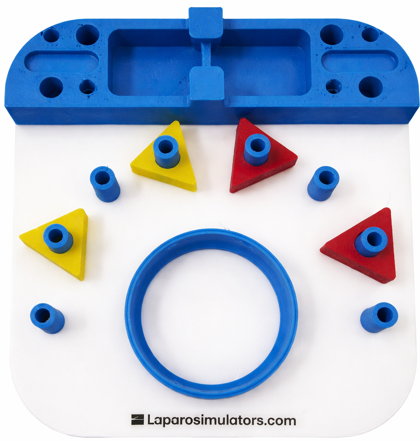}%
  }
  \caption{Central-to-Peripheral Peg Transfer (C2P) task used in VISTA.
  Four perforated triangular objects are transferred from the central
  loading zone \(C_0\) to the target pegs \(P_1\)--\(P_4\) under
  network-impaired video feedback.}
  \label{fig:task_definition}
\end{figure}

As shown in Fig.~\ref{fig:task_definition}, the central circular region
\(C_0\) serves as the loading zone and four peripheral pegs
\(P_1\)--\(P_4\) serve as target locations. At the start of each trial,
four perforated triangular objects are placed in \(C_0\). The operator grasps each object under network-impaired video feedback, transfers it
within the same field of view, aligns its hole with the designated peg,
and releases it. A trial ends when all four objects are placed or when
the predefined time limit of 300\,s is reached.

\begin{figure*}[!t]
  \centering
  \vspace{-2mm}
  \subfloat[PSNR\label{fig:psnr}]{%
    \includegraphics[width=0.285\textwidth]{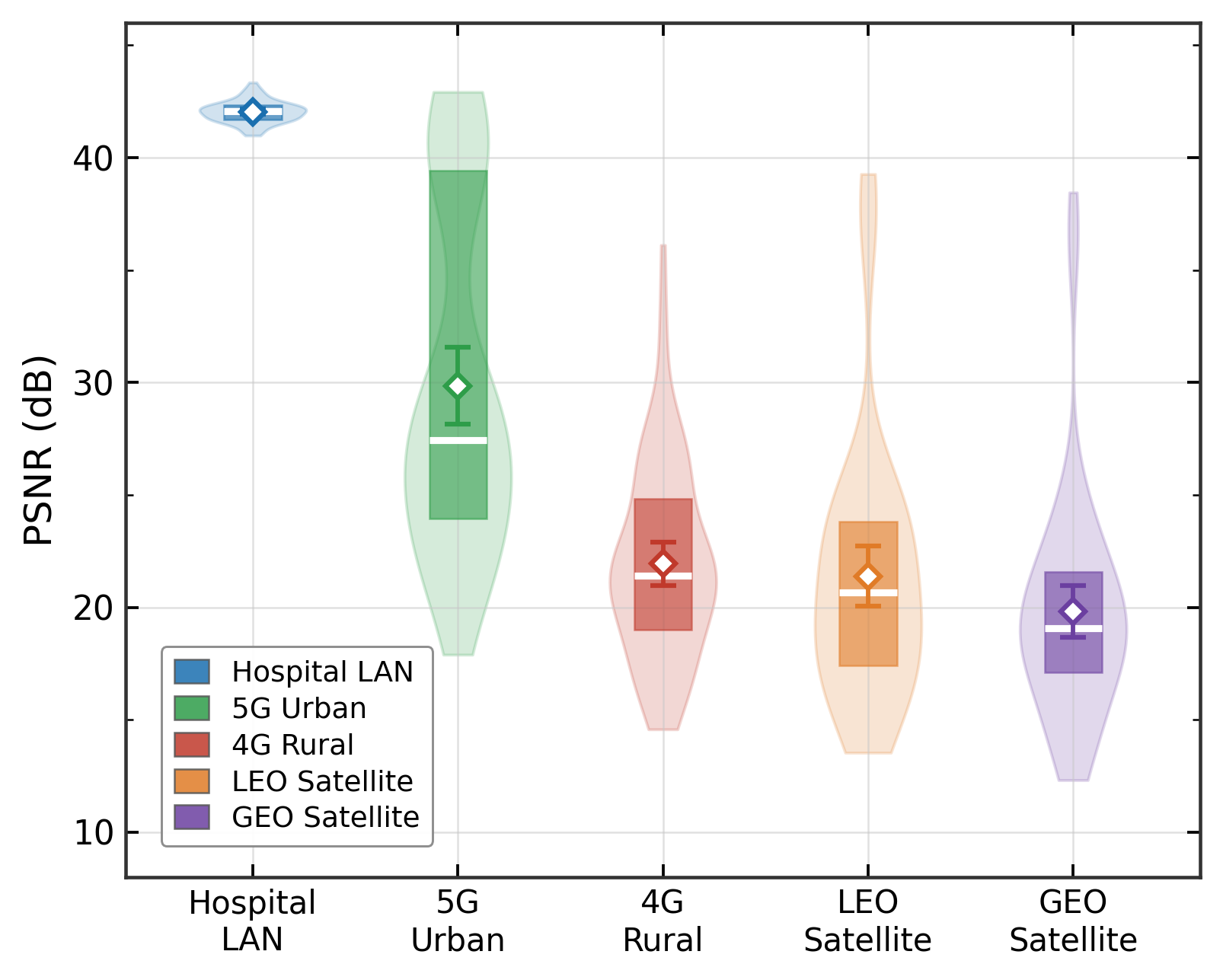}}%
  \hfil
  \subfloat[SSIM\label{fig:ssim}]{%
    \includegraphics[width=0.285\textwidth]{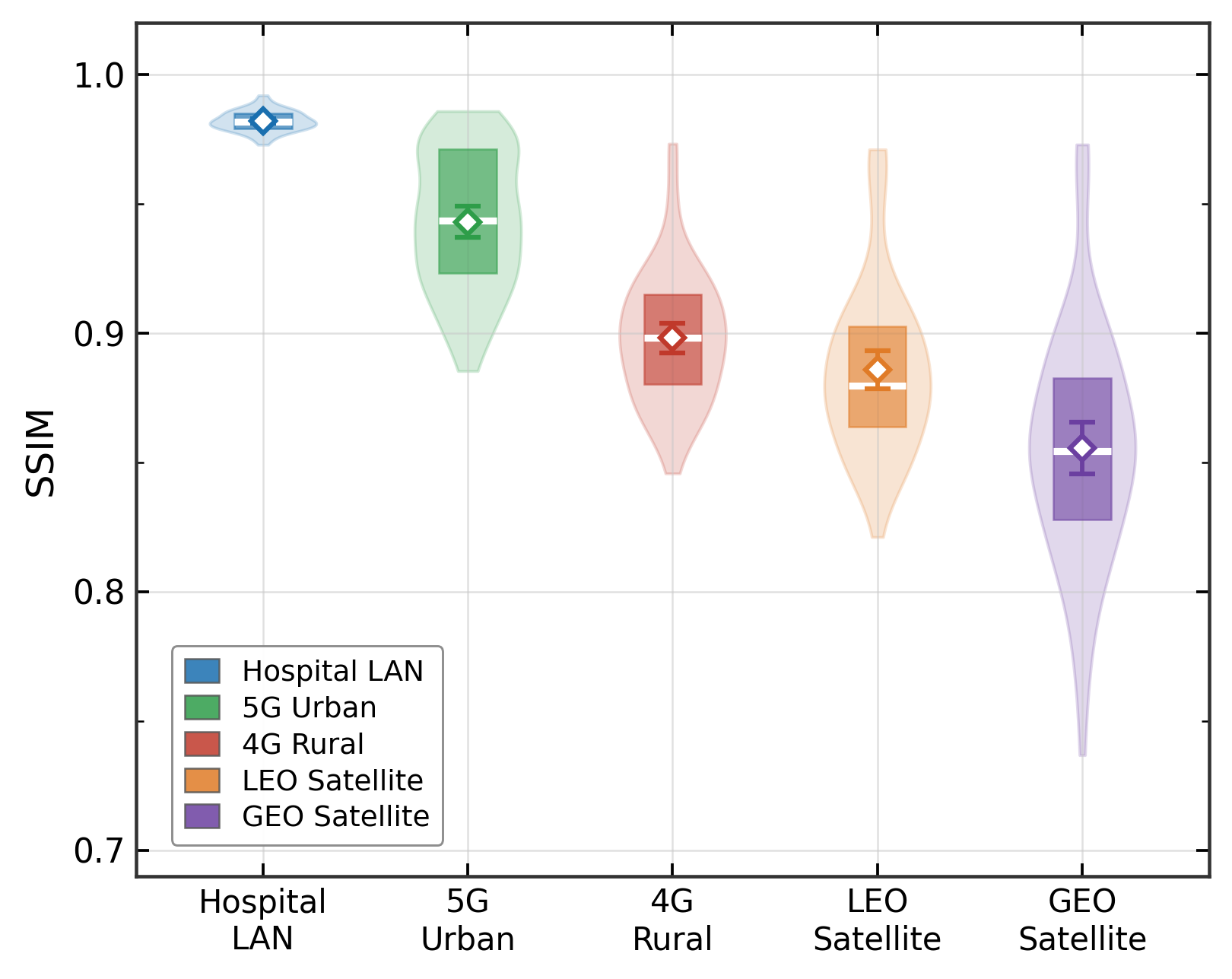}}%
  \hfil
  \subfloat[VMAF\label{fig:vmaf}]{%
    \includegraphics[width=0.285\textwidth]{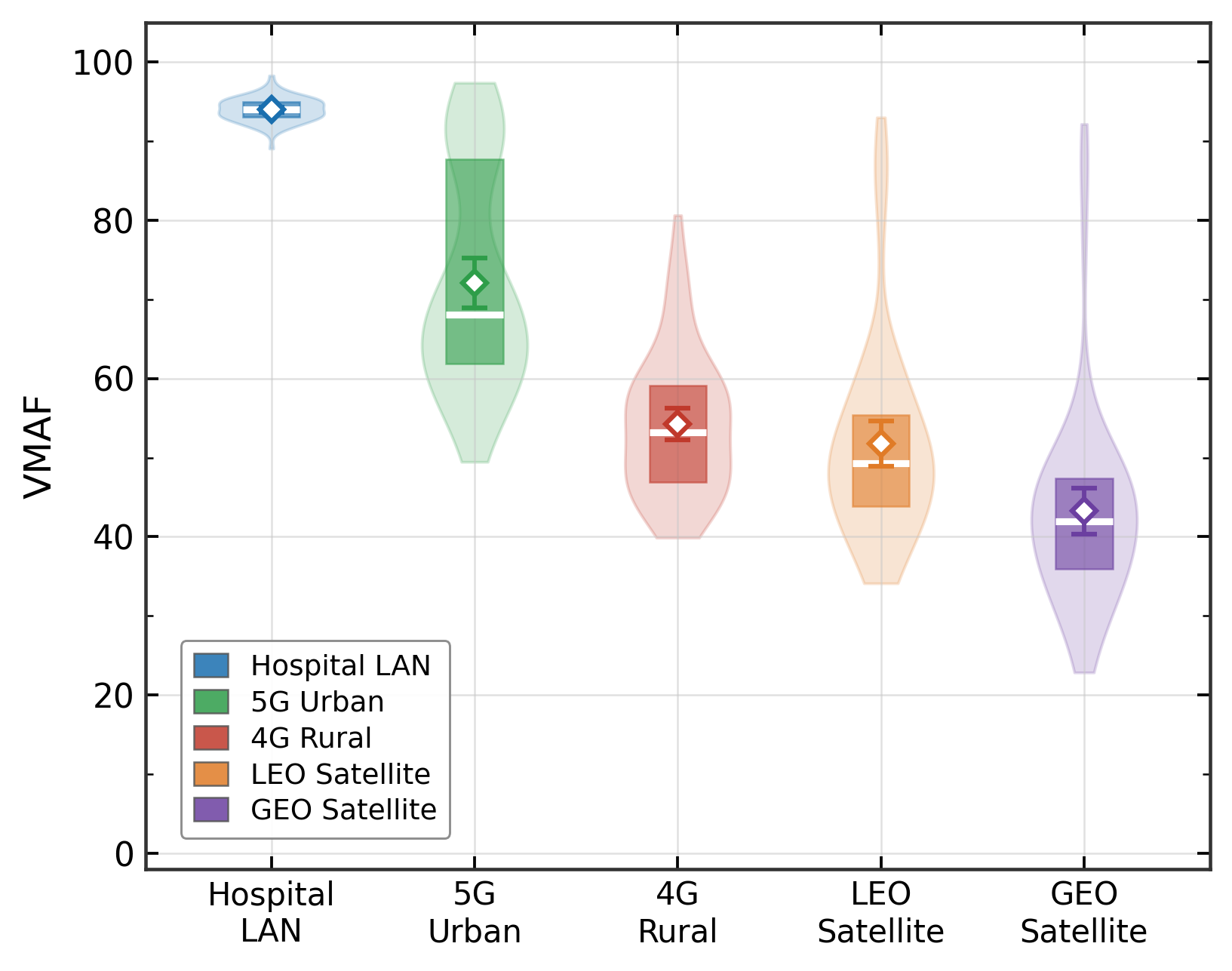}}%
  \\[-10pt]
  \hfil
  \subfloat[Task Completion Time\label{fig:ct}]{%
    \includegraphics[width=0.285\textwidth]{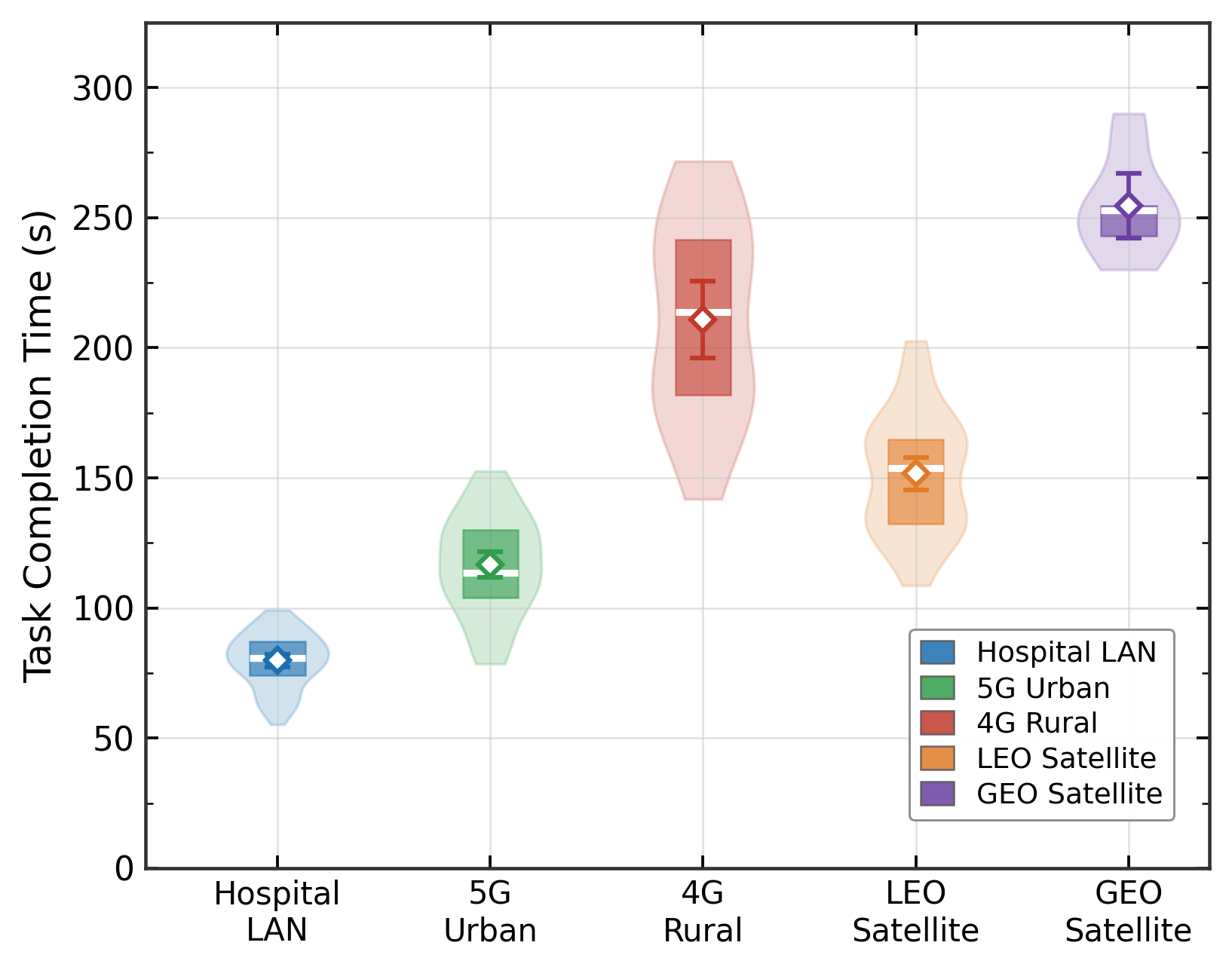}}%
  \hfil
  \subfloat[Trial Outcome Distribution\label{fig:outcome}]{%
    \includegraphics[width=0.285\textwidth]{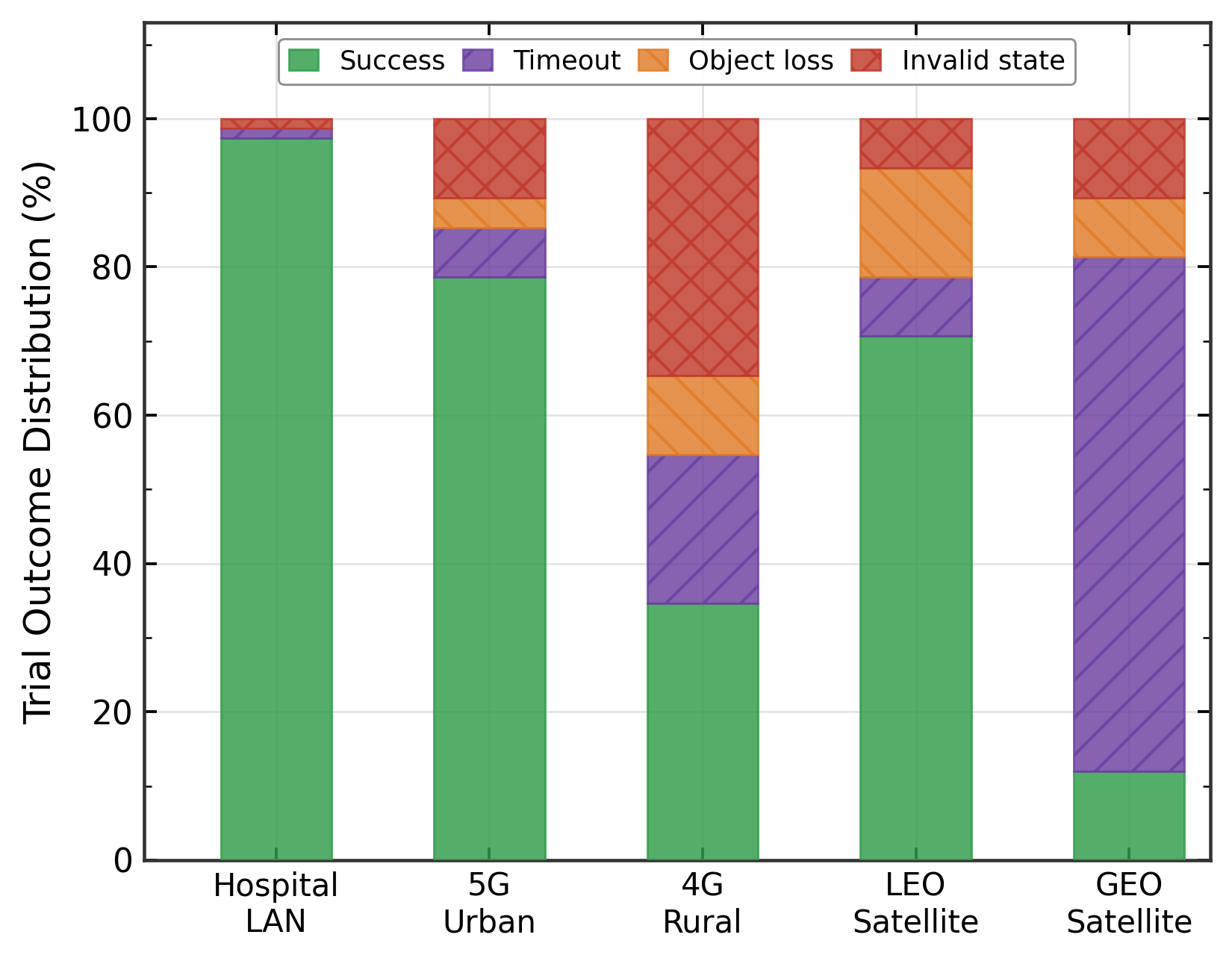}}%
  \hfil
  \vspace{-1mm}
  \caption{Results across the five benchmark tiers.
  Panels~(a)--(c) show objective video quality metrics, where higher
  values indicate better received video quality. Panel~(d) shows task
  completion time for successful trials only, where lower values indicate
  better task efficiency. Panel~(e) shows the distribution of trial
  outcomes, where a higher success proportion and lower failure
  proportions indicate better teleoperation performance.
  In panels~(a)--(d), the coloured box denotes the interquartile range,
  the white line denotes the median, and the diamond denotes the
  mean~$\pm$~95\,\%\,CI. Numerical summary is given in
  Table~\ref{tab:results_summary}.}
  \label{fig:all_metrics}
  \vspace{-1mm}
\end{figure*}

To reduce learning effects, participants completed standardised
pre-experimental training in the Marion Surgical Robot
Simulator.\footnote{Marion Surgical Robot Simulator: \url{https://store.steampowered.com/app/3457580/Marion_Surgical_Robot_Simulator/}}
The simulator was used as a proficiency filter rather than a surrogate
for surgical competence~\cite{moglia2016systematic}. Training metrics
included Grasper Collisions, Distance Travelled, and Offscreen Time.
Operators qualified for the study only after meeting the predefined
criterion in three consecutive sessions: zero grasper collisions and
inter-session variation in travelled distance below $\pm 5\%$~\cite{hogle2008does}.
The study received institutional ethics approval, and all participants
provided informed consent.

A total of 15 qualified operators completed 375 trials. Each operator performed 5 trials per condition across all five tiers
under fixed codec settings, yielding 75 trials per tier. Condition
order was counterbalanced across operators, with short rest periods
between blocks.

\subsection{Video Transport and Receiver Configuration}

The video source is the integrated endoscopic camera of the ASPIRE
Surghero trainer, producing \(640\times480\) RGB video at 30\,fps over
USB/V4L2. H.264 (libx264) is used as the sole codec across all tiers~\cite{x264},
with a Group of Pictures (GOP) size of 15 frames and
tune=zerolatency, reflecting predictive coding in practical
teleoperation video systems~\cite{wang2025influence}. The
tune=zerolatency configuration produces one access unit per
frame, while the fixed GOP limits error propagation to at most 15
frames. Each encoded access unit is packetized into User Datagram Protocol (UDP) datagrams, each carrying up to 1200 bytes of encoded video payload. Each datagram
includes a 16-byte custom header containing the frame ID, chunk index,
total chunk count, and a 64-bit capture timestamp. The payload size is
chosen to remain within the path maximum transmission unit (MTU) and
thereby avoid IP fragmentation. Because the sender and receiver run on the
same workstation, clock synchronization uncertainty is removed from
one-way timestamping, and an adaptive reassembly buffer is used at the
receiver to mitigate false frame drops under high jitter. On H.264
decode failure, the most recent valid frame is redisplayed and the
event is logged as a freeze; concealed frames are excluded from PSNR,
SSIM, and VMAF calculations. Trial timing starts on receipt of the
first decoded frame, yielding an alignment offset bounded by one frame
period. Keeping codec settings constant across all tiers ensures that
quality variation is attributable to network impairment rather than
encoder adaptation. For objective video quality assessment, the source
and received videos were automatically recorded in each trial, and PSNR,
SSIM, and VMAF were computed offline using the Video Quality Measurement
Tool.\footnote{Video Quality Measurement Tool (VQMT): \url{https://videoprocessing.ai/vqmt/}}

\begin{figure}[!t]
  \centering
  \vspace{1mm}
  \includegraphics[
    width=0.90\linewidth
  ]{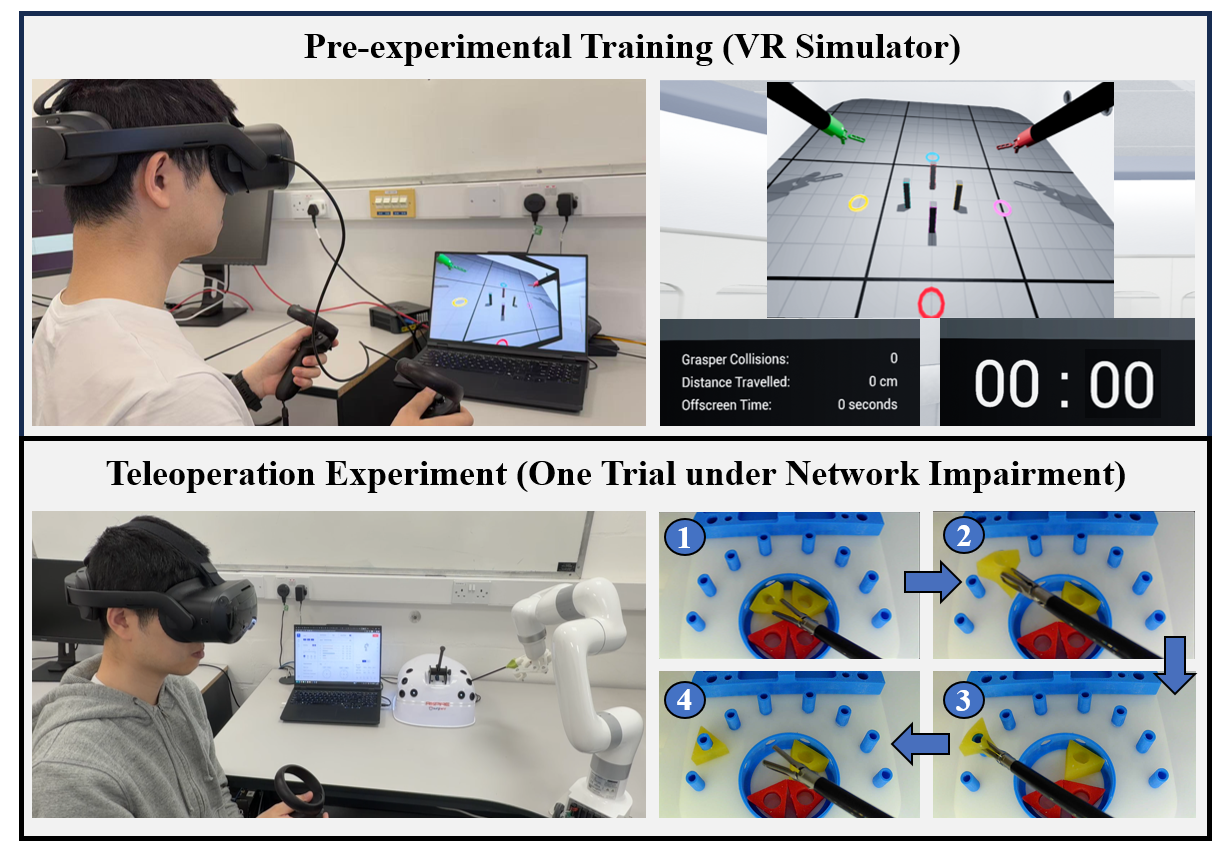}
  \vspace{-2mm}
  \caption{Experimental protocol used in VISTA. Operators first undergo
  pre-experimental training in a VR simulator. After meeting the predefined
  proficiency criteria, they proceed to the teleoperation experiment, where
  one trial consists of a peg transfer task performed under network-impaired
  video feedback.}
  \label{fig:protocol_overview}
  \vspace{-1mm}
\end{figure}

\section{RESULTS AND ANALYSIS}\label{sec:results}

\begin{table*}[!t]
  \centering
  \caption{Key results across the five benchmark tiers.}
  \label{tab:results_summary}
  \renewcommand{\arraystretch}{1.3}
  \setlength{\tabcolsep}{7pt}
  \resizebox{0.98\textwidth}{!}{%
  \begin{tabular}{|l|c|c|c|c|c|c|c|}
    \hline
    \textbf{Networks}
      & \textbf{Completion Time (s)}$^\ast$
      & \textbf{Success Rate (\%)}
      & \textbf{Successful Trials}
      & \textbf{PSNR (dB)}
      & \textbf{SSIM}
      & \textbf{VMAF}
      & \textbf{Freeze Rate (events/min)} \\
    \hline
    Hospital LAN  & $80\pm2$    & 97 & 73 & 42.0 & 0.982 & 94 & 0.0 \\
    \hline
    5G Urban      & $117\pm5$   & 79 & 59 & 29.9 & 0.943 & 72 & 1.1 \\
    \hline
    4G Rural      & $211\pm15$  & 35 & 26 & 21.9 & 0.898 & 54 & 4.5 \\
    \hline
    LEO Satellite & $152\pm6$   & 71 & 53 & 21.4 & 0.886 & 52 & 3.4 \\
    \hline
    GEO Satellite & $255\pm12$  & 12 &  9 & 19.8 & 0.856 & 43 & 5.1 \\
    \hline
  \end{tabular}
  }
  \\[4pt]
  \parbox{0.98\textwidth}{\footnotesize
    $^\ast$Completion time is reported as mean\,$\pm$\,95\% CI for
    successful trials only.
  }
\end{table*}

Trial-level metrics are aggregated across the 75 trials in each
tier, and means are reported with 95\% confidence intervals (CIs).
Within each trial, PSNR, SSIM, and VMAF are averaged over
successfully decoded frames. Freeze rate, success rate, failure mode,
and completion time are recorded at the trial level.

\subsection{Overall Performance}
Fig.~\ref{fig:all_metrics} and Table~\ref{tab:results_summary}
show a clear progression from near-ideal operation to severe
teleoperation impairment. Hospital LAN provides the reference
baseline; 5G~Urban remains largely usable, although its bimodal
quality distribution suggests intermittent H.264 degradation.

The three more impaired tiers present a more important result:
similar average frame quality does not imply similar task
performance. 4G~Rural, LEO~Satellite, and GEO~Satellite
reach mean PSNR values of 21.9, 21.4, and 19.8\,dB,
respectively, yet their success rates diverge substantially to
35\%, 71\%, and 12\%, and their mean completion times for
successful trials span from $152\pm6$\,s to $255\pm12$\,s.
This is consistent with prior telesurgery studies showing that
communication constraints affect performance through more than
one pathway, including both latency and video
degradation~\cite{akasaka2022impact,wang2025influence}.

\subsection{Network Impairments and Task Performance}

The results reveal that similar mean frame quality does not
produce similar task outcomes, suggesting that task performance is
shaped by different impairment patterns across tiers.

In 4G~Rural, the primary driver is sustained spatial quality
degradation: low bandwidth (5\,Mbps) combined with bursty packet
loss ($p_E{=}2\%$, $L_B{=}6$\,pkts) reduces mean PSNR to
21.9\,dB and raises the freeze rate to 4.5\,events/min. Under these conditions, operators frequently lose visual contact with
the instrument tip, leading to a success rate of only 35\% and
the highest proportion of failures due to invalid state among all tiers.
The mean completion time for the minority of successful trials
reaches $211\pm15$\,s, reflecting the additional time required
to recover from degraded spatial feedback.

In GEO~Satellite, the dominant mechanism is fundamentally
different: despite a mean PSNR of 19.8\,dB comparable to
4G~Rural, the 600\,ms one-way delay means that visual feedback can lag
the physical scene by more than one second over a full
perception--action cycle. This persistent temporal
mismatch between the displayed and actual instrument position
causes the operator to act on stale imagery, producing the lowest
success rate across all tiers (12\%) and the longest mean
completion time ($255\pm12$\,s). Timeout is the dominant outcome
(69\%), consistent with operators repeatedly correcting
positional errors induced by outdated visual feedback rather than
failing to perceive the instrument entirely. The one-way delay of
600\,ms substantially exceeds both the 320\,ms clinical ceiling
reported in~\cite{wang2025influence} and the broader latency
limits discussed in telesurgery
studies~\cite{Motiwala2025, Xu2014}.

In LEO~Satellite, mean PSNR (21.4\,dB) and overall success rate
(71\%) suggest moderate impairment, yet the elevated standard
deviation of completion time and the 15\% object-loss rate
indicate that degradation is episodic rather than sustained.
Analysis of individual LEO trials reveals distinct impairment episodes
coinciding with frame loss bursts and PSNR reduction. The most severe episodes reach concealed-frame proportions exceeding 85\%, before recovering to baseline within seconds. These transient episodes disrupt
continuous hand--eye coordination precisely when fine alignment
is required, leading to object loss even when average conditions
remain less severe than in GEO~Satellite or 4G~Rural.

Taken together, these findings show that network impairments
degrade task performance through at least two distinct pathways:
sustained spatial quality loss due to bandwidth limitation and
packet loss (dominant in 4G~Rural), and temporal mismatch between
displayed and actual instrument state caused by high propagation
delay (dominant in GEO~Satellite), with episodic instability
(LEO~Satellite) constituting a third, qualitatively different
mode. Mean frame quality metrics alone cannot distinguish these
pathways, which motivates the multi-metric evaluation pipeline
provided by \textit{VISTA}.

\subsection{Failure Analysis}
The outcome distributions in Fig.~\ref{fig:outcome} show that the
three impaired tiers fail through different mechanisms. In 4G~Rural,
invalid state is the dominant failure mode (35\%), followed by timeout
(20\%), consistent with low spatial quality and frequent freezes
(4.5\,events/min) impairing fine peg alignment. In GEO~Satellite,
timeout dominates (69\%), indicating that operators repeatedly issued
corrections based on stale visual feedback until the 300\,s limit was
reached. In LEO~Satellite, success remains relatively high (71\%), but
object loss increases to 15\%, suggesting episodic disruption to
hand--eye coordination rather than sustained degradation.

These differences confirm that average frame quality alone cannot
explain teleoperation failure. 4G~Rural and LEO~Satellite have
comparable PSNR values but differ by 36 percentage points in success
rate, whereas GEO~Satellite and 4G~Rural share similarly degraded frame
quality but exhibit different dominant failure modes. Spatial fidelity,
propagation delay, and temporal impairment patterns therefore need to
be evaluated jointly in teleoperation safety assessment.

\section{CONCLUSION}
In this paper, we presented \textit{VISTA}, a hardware-in-the-loop benchmark
for evaluating video feedback under network impairments in
robot-assisted laparoscopic teleoperation. By isolating the forward
video path and emulating bandwidth, latency, jitter, and burst loss
in a controlled manner, \textit{VISTA} enables reproducible evaluation of
received video quality, temporal continuity, and task performance.
Results from 375 trials across five representative network tiers
showed that similar average frame quality does not necessarily lead
to similar teleoperation outcomes. \textit{VISTA} therefore provides a
practical basis for evaluating how spatial quality, one-way delay,
and the temporal pattern of network impairment jointly affect
teleoperation performance.

\addtolength{\textheight}{-12cm}   



\begin{thebibliography}{99}

\bibitem{Meara2015}
J.~G. Meara, A.~J.~M. Leather, L.~Hagander, B.~C. Alkire,
N.~Alonso, E.~A. Ameh, S.~W. Bickler, L.~Conteh, A.~J. Dare,
J.~Davies, \emph{et al.},
``Global Surgery 2030: evidence and solutions for achieving
health, welfare, and economic development,''
\emph{Lancet}, vol.~386, no.~9993, pp.~569--624, 2015.

\bibitem{marescaux2001transatlantic}
J. Marescaux, J. Leroy, M. Gagner, F. Rubino, D. Mutter, M. Vix, S. E. Butner, and M. K. Smith,
“Transatlantic robot-assisted telesurgery,”
\textit{Nature}, vol. 413, no. 6854, pp. 379--380, 2001.

\bibitem{rocco2024insights}
B. Rocco, M. C. Moschovas, S. Saikali, G. Gaia, V. Patel, and M. C. Sighinolfi,
“Insights from telesurgery expert conference on recent clinical experience and current status of remote surgery,”
\textit{Journal of Robotic Surgery}, vol. 18, no. 1, p. 240, 2024.

\bibitem{akasaka2022impact}
H. Akasaka, K. Hakamada, H. Morohashi, T. Kanno, K. Kawashima, Y. Ebihara, E. Oki, S. Hirano, and M. Mori,
“Impact of the suboptimal communication network environment on telerobotic surgery performance and surgeon fatigue,”
\textit{PLoS One}, vol. 17, no. 6, p. e0270039, 2022.

\bibitem{3GPP22263}
3GPP,
``Service requirements for Video, Imaging and Audio for Professional Applications (VIAPA),''
\textit{3GPP TS 22.263}, Release 19, 2025.




\bibitem{wang2025influence}
Y. Wang, Q. Ai, T. Shi, B. Gao, W. Zhao, C. Jiang, G. Liu, L. Zhang, H. Li, et al.,
“Influence of network latency and bandwidth on robot-assisted laparoscopic telesurgery: A pre-clinical experiment,”
\textit{Chinese Medical Journal}, vol. 138, no. 3, pp. 325--331, 2025.

\bibitem{Motiwala2025}
Z. Y. Motiwala, A. Desai, R. Bisht, S. Lathkar, S. Misra, and D. D. Carbin,
“Telesurgery: current status and strategies for latency reduction,”
\textit{J. Robot. Surg.}, vol. 19, no. 1, p. 153, 2025.

\bibitem{Baltaci2022}
A. Baltaci, H. Cech, N. Mohan, F. Geyer, V. Bajpai, J. Ott, and D. Schupke,
“Analyzing Real-time Video Delivery over Cellular Networks for Remote Piloting Aerial Vehicles,”
in \textit{Proc. ACM Internet Measurement Conference (IMC)}, 2022, pp. 98--112.

\bibitem{Deng2025TeleSim}
Z. Deng, Z. Yuan, and L. Zou,
“TeleSim: A Network-Aware Testbed and Benchmark Dataset for Telerobotic Applications,”
\textit{arXiv preprint arXiv:2507.04425}, 2025.

\bibitem{Thompson1999}
J.~M. Thompson, M.~P. Ottensmeyer, and T.~B. Sheridan,
``Human factors in telesurgery: Effects of time delay and asynchrony
in video and control feedback with local manipulative assistance,''
\emph{Telemedicine Journal}, vol.~5, no.~2, pp.~129--137, 1999.

\bibitem{Xu2014}
S. Xu, M. Perez, K. Yang, C. Perrenot, J. Felblinger, and J. Hubert,
“Determination of the latency effects on surgical performance and the acceptable latency levels in telesurgery using the dV-Trainer® simulator,”
\textit{Surg. Endosc.}, vol. 28, no. 9, pp. 2569--2576, Sep. 2014, doi: 10.1007/s00464-014-3504-z.

\bibitem{Kumcu2017}
A. Kumcu, L. Vermeulen, S. A. Elprama, P. Duysburgh, L. Platiša, Y. Van Nieuwenhove, N. Van De Winkel, A. Jacobs, J. Van Looy, and W. Philips,
“Effect of video lag on laparoscopic surgery: correlation between performance and usability at low latencies,”
\textit{Int. J. Med. Robot. Comput. Assist. Surg.}, vol. 13, no. 2, Art. no. e1758, 2017, doi: 10.1002/rcs.1758.




\bibitem{tcnetem}
F. Ludovici and H. P. Pfeifer, “tc-netem(8) — Linux manual page,”
\textit{man7.org}. [Online]. Available: \url{https://www.man7.org/linux/man-pages/man8/tc-netem.8.html}. Accessed: Mar. 14, 2026.

\bibitem{gilbert1960}
E.~N. Gilbert,
``Capacity of a burst-noise channel,''
\textit{Bell Syst. Tech. J.}, vol.~39, no.~5, pp.~1253--1265, 1960.

\bibitem{hasslinger2008}
G.~Ha\ss{}linger and O.~Hohlfeld,
``The Gilbert--Elliott model for packet loss in real-time services
on the Internet,''
in \textit{Proc. 14th GI/ITG Conf. Meas., Model. Eval. Comput.
Commun. Syst.\ (MMB)}, Dortmund, Germany, 2008, pp.~1--15.

\bibitem{ITUY1541}
ITU-T,
``Network performance objectives for IP-based services,''
\textit{ITU-T Rec.\ Y.1541}, 2011.

\bibitem{3GPP22261}
3GPP,
``Service requirements for the 5G system,''
\textit{3GPP TS 22.261}, Release 17.

\bibitem{richter2025starlink}
R. Richter, V. Ververis, and V. Bajpai,
``Breaking Through the Clouds: Performance Insights into Starlink's Latency and Packet Loss,''
in \textit{IFIP Networking Conference (Networking)}, 2025.

\bibitem{ITUG114}
ITU-T,
``One-way transmission time,''
\textit{ITU-T Rec.\ G.114}, 2003.

\bibitem{moglia2016systematic}
A. Moglia, V. Ferrari, L. Morelli, M. Ferrari, F. Mosca, and A. Cuschieri,
“A systematic review of virtual reality simulators for robot-assisted surgery,”
\textit{Eur. Urol.}, vol. 69, no. 6, pp. 1065--1080, 2016.



\bibitem{hogle2008does}
N. J. Hogle, W. D. Widmann, A. O. Ude, M. A. Hardy, and D. L. Fowler,
“Does training novices to criteria and does rapid acquisition of skills on laparoscopic simulators have predictive validity or are we just playing video games?,”
\textit{J. Surg. Educ.}, vol. 65, no. 6, pp. 431--435, 2008.

\bibitem{x264}
VideoLAN Organization,
``x264 --- a free H.264/AVC encoder,''
[Online]. Available: \url{https://www.videolan.org/developers/x264.html}.
Accessed: Mar. 14, 2026.






\end{thebibliography}
\end{document}